# Local Minima of a Quadratic Binary Functional with Quasi-Hebbian Connection Matrix

Yakov Karandashev, Boris Kryzhanovsky and Leonid Litinskii

Scientific Research Institute for System Analysis Russian Academy of Sciences (Moscow),

ya_rad_wsem@mail.ru, kryzhanov@mail.ru, litin@mail.ru.

The local minima of a quadratic functional depending on binary variables are discussed. An arbitrary connection matrix can be presented in the form of quasi-Hebbian expansion where each pattern is supplied with its own individual weight. For such matrices statistical physics methods allow one to derive an equation describing local minima of the functional. A model where only one weight differs from other ones is discussed in details. In this case the above-mention equation can be solved analytically. Obtained results are confirmed by computer simulations.

## I. Introduction

Let us examine the problem of minimization of quadratic functional depending on $N$ binary variables $S_i = \{\pm 1\}$, $i = 1, 2, ..., N$:

$$E(\mathbf{S}) = -\frac{1}{N^2} \sum_{i,j=1}^{N} J_{ij} S_i S_j \xrightarrow{\mathbf{S}} \min . \qquad (1)$$

This problem arises in different scientific fields starting with physics of magnetic materials and neural networks up to analysis of results of physical experiments and logistics. The state of the system as a whole is given by $N$-dimensional vector $\mathbf{S} = (S_1, S_2, ..., S_N)$ with binary coordinates $S_i$. These vectors will be called *configuration vectors* or simply *configurations*. The functional $E = E(\mathbf{S})$, which has to be minimized, will be called *the energy* of the state. Without loss of generality the connection matrix $\mathbf{J} = (J_{ij})$ can be considered as a symmetric one ($J_{ij} = J_{ji}$) since the substitution $J_{ij} \to (J_{ij} + J_{ji})/2$ does not change the value of $E$.

In the general case practically nothing is known neither about the number of local minima of the functional (1), nor about their structure. Only in some cases there are more or less clear ideas about the energy surface of the functional. These cases are the Sherrington-Kirkpatrick model of spin glass [1] and high-symmetrical neural network of Hopfield's type [2]. In both cases characters of connection matrices were of great importance.

Statistical physics methods were efficiently used for analysis of the energy surface in the Hopfield model [3]-[5]. In these papers was used the Hebb connection matrix. It is a correlation matrix constructed from a set of $M$ given configurations $\xi^\mu = (\xi_1^\mu, \xi_2^\mu, ..., \xi_N^\mu)$, $\xi_i^\mu = \{\pm 1\}$, $\mu = 1, 2, ..., M$:

$$J_{ij} = \sum_{\mu=1}^{M} \xi_i^\mu \xi_j^\mu .$$

In the theory of neural networks configurations $\{\xi^\mu\}_1^M$ are called *patterns*. We also use this notation. When statistical physics methods are applied, the result is as follows. If patterns $\xi^\mu$ are randomized and independent and their number $M < 0.14 \cdot N$, in the vicinity of each pattern there is necessarily a local minimum of the functional (1). In other words, the neural network works as an associative memory. For the first time this result was obtained in [3], [4].

Up to now it was supposed that statistical physics methods are applicable only for the Hebb connection matrix. However, in the papers [6]-[8] it was shown that any symmetric matrix can be presented as quasi-Hebbian expansion in uncorrelated configurations $\xi^\mu$:

$$J_{ij} = \sum_{\mu=1}^{M} r_\mu \xi_i^\mu \xi_j^\mu . \qquad (2)$$

In Eq. (2) weights $r_\mu$ are determined from the condition of non-correlatedness of configurations $\xi^\mu$, and the number $M$ is determined by the accuracy of approximation of the original matrix $\mathbf{J}$. The representation (2) is called the quasi-Hebbian since all weights $r_\mu$ are different. We hope that this representation would allow one to use statistical physics methods for analysis of an arbitrary connection matrix. First obtained results are presented in this publication.

In Section II with the aid of statistical physics methods the principal equation for a connection matrix of the type (2) is derived. Determination of conditions under which the equation has a solution provides information about local minima of the functional (1). As an example the classical Hopfield model (all the weights are $r_\mu \equiv 1$) is examined in details. In Section III we analyze the case when only one weight differs from the others, which are identically equal: $r_1 \neq r_2 = r_3 = ... = r_M = 1$. This model case can be analyzed analytically up to the end. It was found that a single weight that differs from 1 substantially affects the properties of local minima. Computer simulations confirm this result. Some conclusions and remarks are given in Section IV.

## II. Principle Equation and Hopfield Model

**1. Principle equation.** Here and in what follows we analyze the functional (1) with the quasi-Hebbian connection matrix (2). Let $\mathbf{S} = (S_1, S_2, ..., S_N)$, where $S_i = \pm 1$ define a state of the system. The energy of the state can be presented in the form

$$E(\mathbf{S}) = -\sum_{\mu=1}^{M} r_\mu \left( m_\mu^2(\mathbf{S}) - \frac{1}{N} \right), \tag{3}$$

where $m_\mu(\mathbf{S})$ is the overlap of the state $\mathbf{S}$ with the pattern $\xi^\mu$:

$$m_\mu(\mathbf{S}) = \frac{1}{N} \sum_{i=1}^{N} S_i \xi_i^\mu.$$

Statistical physics methods allow one to obtain equations for the overlap of *a local minimum* of the functional (1) with the patterns $\xi^\mu$. After solving these equations it is possible to understand under which conditions the overlap of a local minimum with the $k$-th pattern is of the order of 1 ($m_k \sim 1$). In other words: under which conditions the local minimum coincides (or nearly coincides) with the $k$-th pattern.

Supposing the dimensionality of the problem to be very large ($N \gg 1$), let us set that the number of patterns $M$ is proportional to $N$: $M = \alpha \cdot N$. In the theory of neural networks the coefficient of proportionality $\alpha$ is called *the load parameter*. Let us suppose that in the vicinity of the pattern $\xi^k$ there is a local minimum of the functional. Repeating calculations performed in [3]-[5] for the Hopfield model, we obtain the system of equations

$$m_k = \mathrm{erf}\left( \frac{r_k m_k}{\sqrt{2}\sigma} \right),$$

$$\sigma^2 = \frac{1}{N} \sum_{\mu \neq k}^{M} \frac{r_\mu^2}{(1 - C \cdot r_\mu)^2}, \tag{4}$$

$$C = \frac{1}{\sigma} \sqrt{\frac{2}{\pi}} \exp\left[ -\left( \frac{r_k m_k}{\sqrt{2}\sigma} \right)^2 \right].$$

Here $m_k$ is the overlap of the local minimum with the $k$-th pattern. In the case of equal weights ($r_\mu \equiv 1$) this system reduces to the well-known system for the Hopfield model (see Eqs. (2.71)-(2.73) in [5]).

Let us introduce an auxiliary variable $y = r_k m_k / \sqrt{2}\sigma$. Excluding $\sigma$ and $C$ from the system (4), we obtain the principle equation

$$\frac{1}{\alpha} = \frac{1}{\gamma^2} \cdot \frac{1}{M} \sum_{\mu \neq k}^{M} \frac{r_\mu^2}{(r_k \cdot \varphi - r_\mu)^2} \tag{5}$$

where

$$\gamma = \sqrt{\frac{2}{\pi}} e^{-y^2} \quad \text{and} \quad \varphi = \frac{\sqrt{\pi}}{2} \frac{\mathrm{erf}\, y}{y} e^{y^2}. \tag{6}$$

The function $\gamma = \gamma(y)$ decreases monotonically, and the function $\varphi = \varphi(y)$ increases monotonically from its minimal value $\varphi(0) = 1$. In what follows these functions are frequently used. For simplicity sometimes we omit their arguments, and use the notations $\gamma$ and $\varphi$ for these functions.

Let us fix the values of external parameters: $N \gg 1$, $\alpha = M/N$, $\{r_\mu\}_1^M$ and $k$. If $y_0$ is a solution of Eq. (5), the overlap of the local minimum with the $k$-th pattern is equal to

$$m_k = \mathrm{erf}\, y_0. \tag{7}$$

With the aid of $y_0$ the value of one other important characteristic $\sigma_k^2 = \sum_{\mu \neq k}^{M} r_\mu^2 m_\mu^2$ can be calculated. Roughly speaking, $\sigma_k^2$ is the weighted sum of squared overlaps of the local minimum from the vicinity of $\xi^k$ with all patterns except the $k$-th one. The expression for $\sigma_k$ has the form

$$\sigma_k = r_k \frac{\operatorname{erf} y_0}{\sqrt{2} y_0}. \tag{8}$$

An important role plays the analysis of conditions under which Eq. (5) can be solved. For example, it is useful to determine the critical value of the load parameter $\alpha_c$ for which the solution of Eq. (5) still exists, but for $\alpha$ that are larger than $\alpha_c$ there is no solution of Eq.(5). All characteristics corresponding to $\alpha_c$ are marked off with the same subscript, namely they are $y_c$, $m_c$ and $\sigma_c$.

**2. The Hopfield model:** $r_\mu \equiv 1$. Since all patterns are equivalent, in Eqs. (4), (5) the subscript $k$ can be omitted. It is easy to see that in this case Eq. (5) has the form

$$\alpha = \gamma^2 (\varphi - 1)^2 \tag{9}$$

The plot of right-hand side of Eq. (9) is shown in Fig. 1. We see that for all $\alpha$ less than a critical value $\alpha_c$, there are two solutions: $\bar{y}_\alpha$ and $y_\alpha$ ($\bar{y}_\alpha < y_\alpha$). We are interested in the larger solution $y_\alpha$. The solution $\bar{y}_\alpha$ is a spurious one, and it has to be omitted.

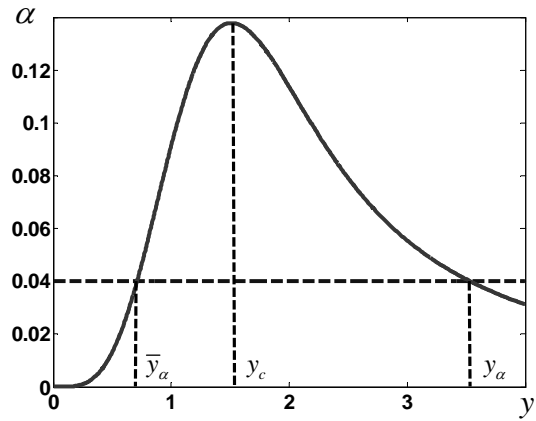

**Fig. 1.** Graphical solution of Eq. (9) for the Hopfield model is shown. Solid line is the graph of the function in the right-hand side of Eq.(9). Dashed line corresponds to the value of load parameter $\alpha = 0.04$.

In general, when Eq. (5) has some solutions, the solution with maximal overlap $m$ corresponds to the minimal value of the free energy. Consequently, it is necessary to pick out the solution that is located as much as possible to the right on the abscissa axis. In what follows we use this rule.

Let us return to our analysis of the plot in Fig.1. When $\alpha$ increases the solution $y_\alpha$ shifts to the left. In the same time the overlap of the local minimum with the pattern also decreases. In other words, the local minimum little by little moves away from the pattern. Increasing $\alpha$ we reach the critical value $\alpha_c$. The critical value $\alpha_c$ is determined by maximum of the right-hand side of Eq. (9). It is not difficult to calculate its value: $\alpha_c \approx 0.138$. This well-known result related to the critical value of the load parameter firstly was obtained in [3], [4].

The expression in the right-hand side of Eq. (9) reaches its maximum value at the point $y_c \approx 1.511$. Equation (7) allows one to calculate the critical value of the overlap of the local minimum with the pattern: $m_c \approx 0.967$. We see that even for the maximal load parameter $\alpha_c$ the local minimum is very close to the pattern. If the load parameter $\alpha$ is less than $\alpha_c$, the overlap of the local minimum with the pattern even closer to 1. Since in this case an arbitrary pattern is considered, we can say that while $\alpha < \alpha_c$ there is a local minimum of the functional (1) in the vicinity of each pattern. It can be shown that the depths of all these local minima are approximately the same [5].

What happens when the parameter $\alpha$ exceeds the critical value $\alpha_c$? In this case there is no solution of Eq. (9). As they say, "a breakdown" of the solution happens. Detailed analysis [3]-[5] shows that the overlap decreases abruptly almost to zero value. In terms of statistical physics this means that in our system the first kind transition takes place. The energy surface of the functional (1) changes: local minima in vicinities of patterns cease to exist.

### III. The Case of One Differing Pattern

Interesting is the case when all weights $r_\mu$ except one are equal to 1, and only one weight differs from other. Without loss of generality we can write

$$r_1 = \tau, \quad r_2 = r_3 = ... = r_M = 1. \tag{10}$$

The first weight can be both larger and less than one.

It might seem that the difference from the classical Hopfield model has to be small: enormous number of patterns with the same weight takes part in the formation of the connection matrix and only one pattern provides a different contribution. Intuition suggests that the influence of a single pattern with the individual weight $\tau$ has to be negligible small comparing with contribution of infinitely large number of patterns with the same weight $r_\mu = 1$. However, this is not the case. One pattern with an individual weight $\tau$ can substantially affect the distribution of local minima. Let us examine separately what happens with local minimum in the vicinity of the pattern with the individual weight $\tau$, and what happen with local minima near other patterns whose weights are equal to 1.

**1. Pattern with an individual weight:** $r_1 = \tau$. For this pattern equation (5) has the form

$$\alpha = \gamma^2 (\tau \cdot \varphi - 1)^2. \tag{11}$$

If $y$ is a solution of Eq. (11), the overlap of the local minimum with the first pattern is $m^{(1)} = erf(y)$. The superscript "(1)" emphasizes that we deal with the overlap of the local minimum with the pattern number 1.

The point of breakdown $y_c^{(1)}(\tau)$, where the right-hand side of Eq.(11) reaches its maximum, is the solution of the equation

$$\varphi(y) = 1 + \frac{2y^2}{\tau}. \tag{12}$$

After finding $y_c^{(1)}(\tau)$ the critical characteristics $m_c^{(1)} = erf(y_c^{(1)})$ and $\alpha_c^{(1)}$ can be calculated. It turns out that for $\tau > 1$ Eq.(12) has a nontrivial solution only if $\tau \le 3$. When $\tau$ is larger than 3, this equation has only trivial solution: $\tau > 3 \Rightarrow y_c^{(1)}(\tau) \equiv 0$. This means that the overlap of the local minimum with the pattern vanishes: $m_c^{(1)} = erf(y_c^{(1)}) \equiv 0$. So, when $\tau > 3$ in the system the phase transition disappears. This is an unexpected result. We do not understand the reason why the phase transition disappears. (We have for $\tau \ge 3$: $\sigma_c^{(1)} = \tau \sqrt{\frac{2}{\pi}}$ and $\alpha_c^{(1)} = 2(\tau-1)^2/\pi$.)

In Fig.2a we present graphs of right-hand side of Eq. (11) for three different values of the weight coefficient: $\tau = 1$, $\tau = 2$ and $\tau = 3$. The case of $\tau = 1$ corresponds to the classical Hopfield model. We see that increasing of the coefficient $\tau$ above 1 on the one hand is accompanied by an increase of the critical value $\alpha_c^{(1)}(\tau)$, and on the other hand it is accompanied by steady removal of the breakpoint $y_c^{(1)}(\tau)$ toward 0. As a result, when $\tau$ increases the overlap of the local minimum with the pattern $m_c^{(1)}(\tau)$ in the critical point decreases.

Generally speaking, the decrease of the overlap, accompanying an increase of the weight coefficient $\tau$, is very unusual. Let us point out that this behavior relates only to the critical values $y_c^{(1)}(\tau)$ and $m_c^{(1)}(\tau)$, i.e. to their values at the breakdown point. Absolutely other situation takes place if $\alpha < \alpha_c$.

Indeed, Fig.2a shows intersections between a straight line parallel to abscissa axis and the graphs representing the right-hand side of Eq. (11) for different values of $\tau$. Abscissa values of each intersection are solutions of the equation. We see that when $\tau$ increases the point of intersection shifts to the right, i.e. in the direction of its greater values. This means that when $\tau$ increases the overlap $m^{(1)}(\tau)$ of the local minimum with the pattern also increases. This behavior of the overlap is in agreement with the common sense: the greater the weight of the pattern $\tau$, the greater its influence. Then the overlap of this pattern with the local minimum has to be greater.

Now let us examine the interval [0,1] of the weight $\tau$. Analysis of Eqs. (11), (12) shows that when $\tau$ decreases from 1 to 0 the maximum point $y_c^{(1)}(\tau)$ steadily shifts to the right, and the maximum value $\alpha_c^{(1)}$ steadily decreases (see the behavior of maxima of the curves in Fig.2b). In other words, when the weight coefficient $\tau$ decreases the critical value of the overlap $m_c^{(1)}(\tau)$ tends to 1. Note, when $\tau < 1$ the function $\tau \cdot \varphi(y) - 1$ vanishes necessarily. By $y_0(\tau)$ we denote the point where this function equals zero. The behavior of the curve to the left from $y_0(\tau)$ does not interesting for us, since in this region Eq. (11) has only spurious solutions.

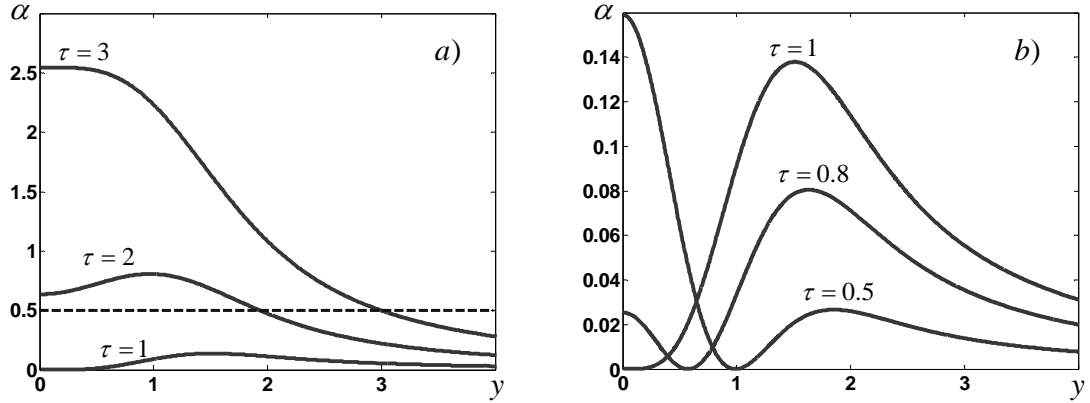

**Fig. 2.** The graphs of the right-hand side of Eq. (11) for different values of $\tau$. *a)* $\tau \geq 1$; the dashed straight line intersects the graphs corresponding to different values of $\tau$; *b)* $\tau \leq 1$; minima of each graph are equal to zero.

Graphs showing the behavior of critical characteristics $m_c^{(1)}(\tau)$ and $\alpha_c^{(1)}(\tau)$ are presented on two left panels of Fig.3.

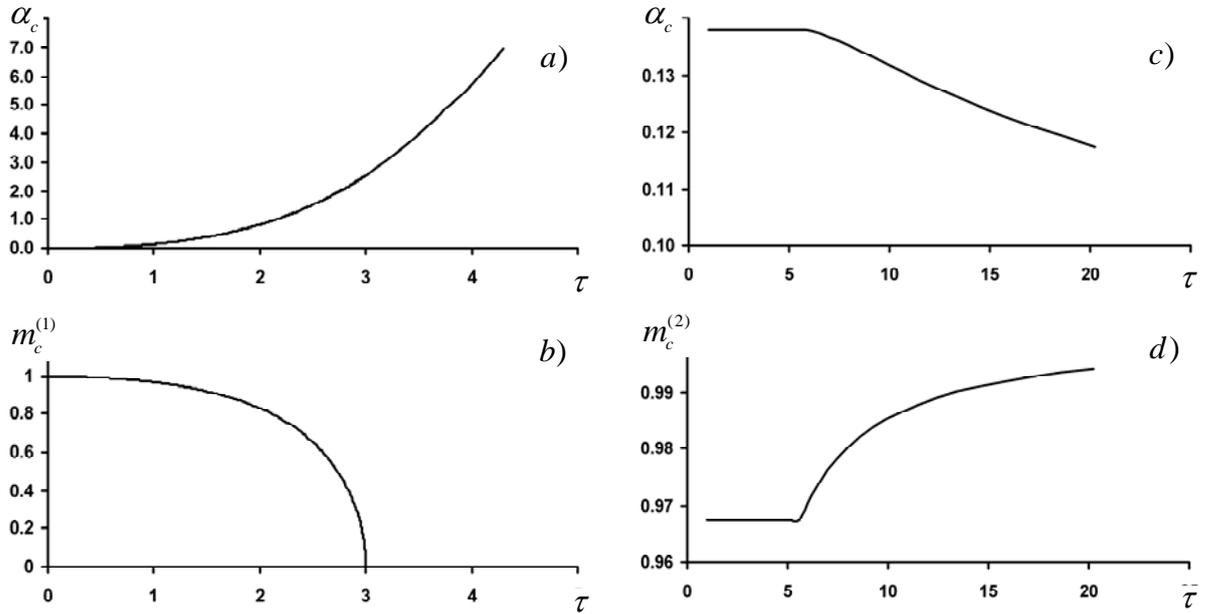

**Fig. 3.** Critical values of the load parameter $\alpha_c$ (two upper panels) and overlap of the local minimum with the pattern $m_c$ (two lower panels) as functions of the weight coefficient $\tau$. The curves on the left panels (*a* and *b*) correspond to the pattern with individual weight $r_1 = \tau$; the curves on the right panels (*c* and *d*) correspond to patterns with the same weight $r_\mu = 1$, $\mu \geq 2$.

**2. Patterns with the same weight:** $r_\mu = 1$, $\mu \geq 2$. Let us examine how the overlap of the local minimum with one of the patterns whose weight coefficient is equal to 1 depends on the value $\tau$. Since all these patterns are equivalent, we choose the pattern with number "2". With the superscript (2) we mark off the characteristics $m^{(2)}, y^{(2)}, \alpha^{(2)}$ that are interesting for us. Now Eq. (5) has the form:

$$L(y) = \alpha, \text{ where } L(y) = \frac{\gamma^2(y)(\varphi(y)-1)^2(\varphi(y)-\tau)^2}{(1-\varepsilon)(\varphi(y)-\tau)^2 + \varepsilon\tau^2(\varphi(y)-1)^2} \text{ and } \varepsilon = \frac{1}{M}. \quad (13)$$

When $M \to \infty$, the quantity $\varepsilon$ tends to zero. However, we cannot simply take $\varepsilon = 0$, since (at least for $\tau > 1$) for some value of $y$ the denominator of $L(y)$ necessarily vanishes. Then it is impossible to cancel identical functions in the numerator and denominator of $L(y)$. So, we analyze Eq. (13) for a small, but finite value of $\varepsilon$ and then we tend it to zero. This way of analysis is correct.

When $\tau \leq 1$, the function $\varphi(y) - \tau$ nowhere vanishes, and we can simply take $\varepsilon = 0$. In this case the expression for $L(y)$ turns into $\gamma^2(y)(\varphi(y)-1)^2$. Then Eq. (13) transforms in Eq. (9) that corresponds to the Hopfield model. Consequently, until the value of $\tau$ belongs to the interval $\tau \in (0,1]$ the following is true:

$$y_c^{(2)}(\tau) \equiv y_c \approx 1.511, \; \alpha_c^{(2)}(\tau) \equiv \alpha_c \approx 0.138, \; m_c^{(2)}(\tau) \equiv m_c \approx 0.967. \quad (14)$$

Let us examine the region $\tau > 1$. In this case the function $\varphi(y) - \tau$ vanishes at the point $y_0(\tau)$. The value of $y_0(\tau)$ is determined by equation:

$$\varphi(y_0(\tau)) = \tau. \quad (15)$$

Out of the small vicinity of the point $y_0(\tau)$ the parameter $\varepsilon$ in Eq. (13) can be tended to zero. At that the expression for the function $L(y)$ is as in the case of the Hopfield model: $L(y) = \gamma^2(y)(\varphi(y)-1)^2$. At the point $y_0(\tau)$ itself for small, but finite value of $\varepsilon$, the function $L(y)$ is equal to zero: $L(y_0(\tau)) = 0$. If $y$ is from the vicinity of the point $y_0(\tau)$ and it tends to $y_0(\tau)$, for any finite value of $\varepsilon$ the curve $L(y)$ quickly drops to zero. Thus, for any finite value of $\varepsilon$ the graph of the function $L(y)$ practically everywhere coincides with the curve $\gamma^2(y)(\varphi(y)-1)^2$ that corresponds to the Hopfield model. And in the vicinity of the point $y_0(\tau)$ the curve $L(y)$ has a narrow dip up to zero whose width is proportional to the value of $\varepsilon$.

As long as the weight $\tau < \varphi(y_c) \approx 5.568$, the point $y_0(\tau)$ is at the left of $y_c$. For this case in Fig. 4a we show the curve $L(y)$. The maximum of the curve $L(y)$ corresponds to the critical point $y_c \approx 1.511$ and it does not depend on $\tau$. Consequently, the equalities (14) are justified not only for $\tau \in [0,1]$, but in the wider interval $0 < \tau \leq \varphi(y_c) \approx 5.568$.

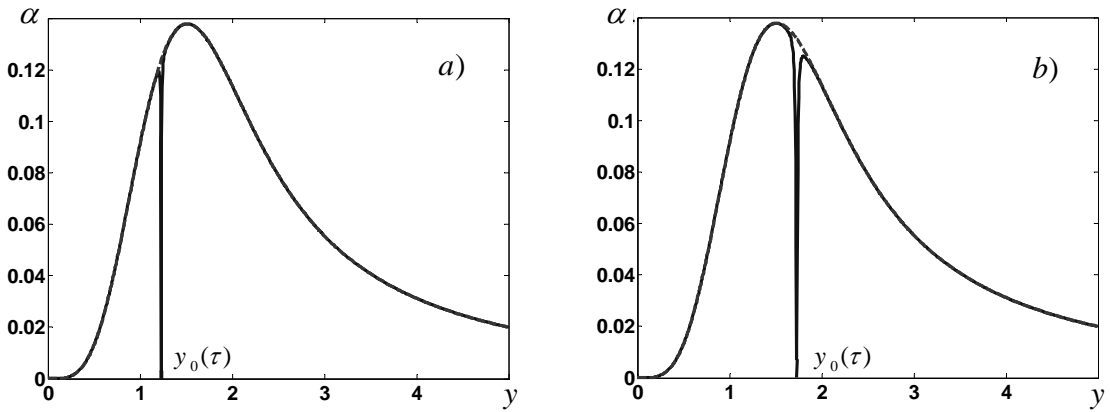

**Fig. 4.** The graph of the function $L(y)$ from Eq. (13), when $\varepsilon = 10^{-5}$ (solid line): a) when $\tau = 3 < \varphi(y_c)$, the point $y_0(\tau)$ is on the left of $y_c \approx 1.511$; b) when $\tau = 10 > \varphi(y_c)$, the point $y_0(\tau)$ is on the right of $y_c \approx 1.511$. Dashed line shows the difference from the classical Hopfield model.

On the contrary, for the values of the weight $\tau > 5.568$ the point $y_0(\tau)$ is on the right of $y_c$. For this case an example of the curve $L(y)$ is shown in Fig. 4*b*. The maximum point of the curve $L(y)$, which is interesting for us, coincides with the peak of the curve that is slightly on the right of the point $y_0(\tau)$. From continuity conditions it is evident that when $\varepsilon \to 0$ this peak shifts to the point $y_0(\tau)$. Consequently, when $\varepsilon \to 0$ for $\tau > 5.568$ we have

$$y_c^{(2)}(\tau) \equiv y_0(\tau) > 1.511, \quad m_c^{(2)}(\tau) \equiv erf(y_c^{(2)}(\tau)) > 0.967, \quad \alpha_c^{(2)}(\tau) = \frac{2}{\pi}(\tau-1)^2 e^{-2y_0^2(\tau)} < \alpha_c = 0.138. \tag{16}$$

Let us summarize the obtained results for patterns with the same weights: $r_\mu = 1$, $\mu \geq 2$. Firstly, as far as the value of $\tau$ is less than $\varphi(y_c) = 5.568$, all characteristics of local minima do not depend on $\tau$ and exactly coincide with characteristics of the Hopfield model. Secondly, as soon as the value of $\tau$ exceeds $\varphi(y_c) = 5.568$ the situation changes. In this case the break-down point of the solution depends on $\tau$ and coincides with $y_0(\tau)$ that is the solution of Eq. (15). Note, when $\tau$ increases the critical load parameter $\alpha_c(\tau)$ decreases and the overlap of the local minimum with the pattern increases and gradually tends to 1 (see Eq. (16)). The graphs showing the behavior of the critical characteristics $m_c^{(2)}(\tau)$ and $\alpha_c^{(2)}(\tau)$ are presented at the two right panels in Fig.3.

**3. Computer simulations.** The obtained results were verified with the aid of computer simulations. For a given value of $N$ the load parameter $\alpha$ was fixed. Then $M = \alpha N$ randomized patterns were generated, and they were used to construct the connection matrix with the aid of Eq. (2). The weights were defined by Eq. (10). When choosing the weight coefficient $\tau$ we proceeded from following reasons. Let $\alpha$ be a fixed value of the load parameter. Then we found the value of the weight $\tau(\alpha)$ for which the given $\alpha$ was a critical one (this can be done when solving Eqs. (11) and (12) simultaneously). We also defined the break-down point of the solution $y_c(\alpha)$. If constructing the connection matrix with the weight $\tau$ that is equal to $\tau(\alpha)$, the mean value of the overlap of the local minimum with the pattern has to be close to $m_c(\alpha) = erf(y_c(\alpha))$. If the weight $\tau$ is less than $\tau(\alpha)$ the mean value of the overlap with the pattern has to be close to 0. If the weight $\tau$ is larger than $\tau(\alpha)$, the mean value of the overlap has to be larger than $m_c(\alpha)$. Under further increase of $\tau$ the mean overlap has to tend to 1.

To verify the theory relating to the pattern with the individual weight $r_1 = \tau$, three experiments has been done. In all experiments the dimensionality of the problem was $N = 10000$. For each load parameter $\alpha$ the mean value of overlap (with the single pattern) $<m>$ was calculated with the aid of ensemble averaging. Ensemble consisted of 10 different matrices. For testing three values of $\alpha$ were chosen. Let us list them together with values of $\tau(\alpha)$ and $m_c(\alpha)$: 1) $\alpha = 0.12$, $\tau(\alpha) \approx 0.944$, $m_c(\alpha) \approx 0.971$; 2) $\alpha = 0.38$, $\tau(\alpha) \approx 1.501$, $m_c(\alpha) \approx 0.919$; 3) $\alpha = 3.0$; for this value of $\alpha$ there is no jump of the overlap, but beginning from $\tau(\alpha) \approx 3.171$, the overlap has to increase smoothly.

In Fig.5 the graphs for all three values of the load parameter are presented. Theoretical characteristics $m_c^{(1)}(\tau)$ are shown by dashed lines. The results of computer simulations are given by solid lines with markers.

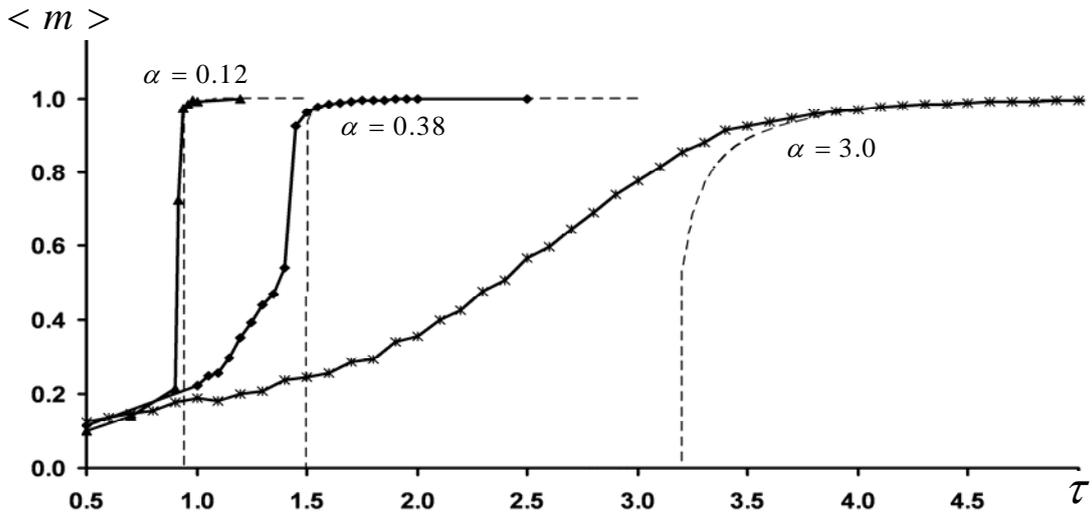

**Fig. 5.** Theory and experiments for the pattern with the individual weight $r_1 = \tau$. The graphs correspond to three different values of the load parameter $\alpha = 0.12, 0.38$ and $3.0$. Solid lines are the results of experiments; dashed lines show theory.

For $\alpha = 0.12$ and $\alpha = 0.38$ on the experimental curves we clearly see expected jumps of mean overlap that have place in the vicinities of critical values of the weights $\tau(\alpha) \approx 0.944$ and $\tau(\alpha) \approx 1.501$, respectively. At the left side the values of the mean overlap are not equal to zero. First, this can be due to not sufficiently large dimensionality of the problem. The point is that all theoretical results are related to the case $N \to \infty$. For our computer simulations we used very large, but finite dimensionality $N$. Second, the theory is correct when the mean overlap is close to 1 (but not to 0). In this region of values of $\tau$ we have rather good agreement of the theory with computer simulations.

For the third load parameter $\alpha = 3.0$ we expected a smooth increasing of the mean overlap $<m>$ from 0 to 1. Indeed, the last right solid curve increases smoothly without any jumps. However, according our theory the increasing ought to start beginning from $\tau \approx 3.1$, while the experimental curve differs from zero much earlier. This discrepancy between our theory and experiment can be explained in the same way as it has been done in the end of the previous paragraph.

To verify our theory relating to patterns with equal weights $r_\mu \equiv 1$, $\mu \geq 2$ we used the same procedure. For the load parameter $\alpha = 0.12$ and several dimensionalities $N$ ($N = 3000$, 10000 and 30000) we calculated the mean overlap $<m>$ of the local minimum with the nearest pattern for different weights $\tau$. We averaged both over $M-1$ patterns of the given matrix and over 10 randomized matrices constructed for the given value of $\tau$.

According to the theory, for $\alpha = 0.12$ breakdown of the overlap $<m>$ has to take place when $\tau \approx 17.1$. In Fig. 6 this place is marked by the right dashed straight line with the label "$M \to \infty$". If $N$ and $M$ really were infinitely large just in this place the breakdown of $<m>$ had to take place. In Fig. 6 the real dependency of $<m>$ on $\tau$ that was observed in our experiments was shown by three solid lines corresponding to different dimensionalities $N$.

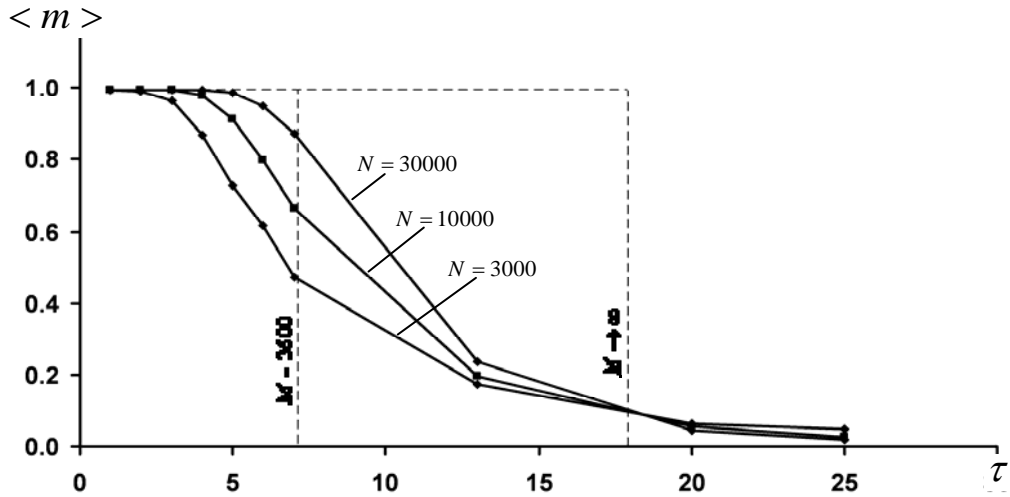

**Fig. 6.** Theory and experiments for patterns with the same weights $r_\mu \equiv 1$ when $\alpha = 0.12$. Solid lines show the results of experiments for three values of dimensionality $N$.

Noticeable difference between the theory and computer simulations must not confuse us. Apparently this difference is due to finite dimensionalities of experimental connection matrices. Earlier computer verifications of classical theoretic results faced just the same problems [10], [11]. As a way out the authors of these papers extrapolated their experimental results into the region of very large dimensionalities $N$.

Note, when $N$ increases the experimental curve in Fig. 6 tends to "theoretical step-function", which is indicated with the aid of dashed line. A correction due to finite dimensionality of the problem can be taken into account if we insert the explicit expression $\varepsilon = 1/M$ in Eq. (13). Then for $N = 30000$ we have $M = \alpha N = 3600$. When this value of $M$ is used in Eq. (13), we obtain that the breakdown of the overlap $<m>$ has to take place not in the vicinity of $\tau \approx 17.1$, but much earlier when $\tau \approx 7.1$. The corresponding dashed line with the label $M = 3600$ is shown in Fig. 6. Its location noticeably better correlates with experimental curves.

**4. Depths of local minima.** Let us find out how the depth of local minimum depends on the value of $\tau$. By the depth of the local minimum we imply the modulus of its energy (3), $|E|$. For different matrices the energies can be compared only if the same scale is used for calculation. We normalized elements of each matrix dividing them by the square root from the dispersion of matrix elements: $\sigma_J = \sqrt{\tau^2 + \alpha N}$ is the standard deviation of quasi-Hebbian matrix elements of the form (2) for the weights (10); mean value of matrix elements is equal to 0. Then for the local minimum depth we obtain the expression

$$|E| = \frac{1}{\sigma_J} \sum_{\mu=1}^{M} r_\mu \left( m_\mu^2 - \frac{1}{N} \right).$$

Let us examine the local minimum in the vicinity of the pattern $\xi^1$ with the weight $r_1 = \tau$. For this minimum we obtain according to Eqs. (5)-(12)

$$|E_1| = \frac{1}{\sigma_J} \left( \tau m_1^2 + \sigma_1^2 - \alpha \right), \quad \sigma_1 = \tau \gamma \varphi. \tag{17}$$

Here $\alpha$ is the load parameter, $m_1$ and $\sigma_1$ are calculated according to Eqs. (7) and (8), $\gamma$ and $\varphi$ are given by the expressions (6). Analysis of Eq. (17) and numerical calculations show that when $\tau$ increases the depth of the minimum increases monotonically. While $\tau$ is less than a critical value $\tau_c(\alpha)$ the local minimum is very far from the pattern ($m_1 \approx 0$) and its depth is equal to zero. When $\tau > \tau_c(\alpha)$, minimum quickly approaches to the pattern ($m_1 \to 1$) and its depth increases in a jump-like way. After that $|E_1|$ increases as $\tau/\sqrt{\tau^2 + \alpha N}$. In Fig. 7 for different load parameters $\alpha$ the dependence of $|E_1|$ on $\tau$ is shown. When $\tau$ becomes very large ($\tau > \sqrt{\alpha N}$) the depth of the minimum asymptotically tends to 1. It could be thought that for such large $\tau$ the matrix $\mathbf{J}$ is constructed with the aid of only one pattern $\xi^1$.

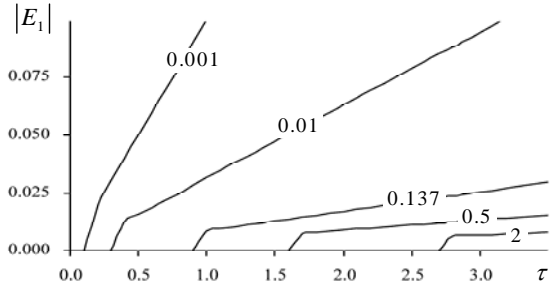
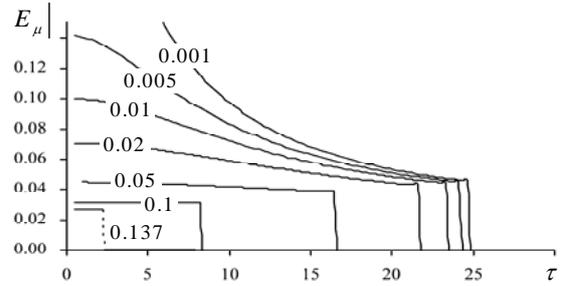

**Fig. 7.** The dependence of the depth of the local minimum from the vicinity of the pattern $\xi^1$ on $\tau$. The load parameter $\alpha$ changes from 0.001 to 2.

**Fig. 8.** The dependence of the depth of the local minimum from the vicinity of one of the patterns $\xi^\mu$ ($r_\mu = 1$) on $\tau$. The load parameter $\alpha$ changes from 0.001 to 0.137.

Now let us examine local minima near patterns with the same weight $r_\mu = 1$ ($\mu \geq 2$). First, independent of the value of the weight $\tau$, these local minima exist only when $\alpha \leq \alpha_c$ (see Eqs. (14), (16)). The expression for the depth of the $\mu$-th minimum has the form:

$$|E_\mu| = \frac{1}{\sigma_J} \left( m_\mu^2 + \sigma_\mu^2 - \alpha \right), \quad \sigma_\mu = \gamma \varphi. \tag{18}$$

Analysis of Eq. (18) shows that when $\tau$ increases, the depth of the local minimum decreases (see Fig. 8). When $\tau$ becomes equal to some critical value $\tau_c(\alpha)$, the depth of the minimum reaches its minimal value $E_c = E_c(\alpha)$:

$$E_c(\alpha) = \frac{2}{\sqrt{\alpha_c N}} - \alpha_c \alpha.$$

For further increase of $\tau$ the overlap $m_\mu$ step-wise becomes equal to zero: minimum abruptly "goes away" from the pattern $\xi^\mu$, and its depth $|E_\mu|$ drops to zero. From the point of view of the associative memory this means that for $\tau > \tau_c(\alpha)$ the memory of the network is destroyed.

## IV. Discussion and Conclusions

Local minima of the functional (1) are fixed points of an associative neural network. In this paper we examined minima localized near patterns, which were used to construct the connection matrix. All obtained results can be interpreted in terms of neural networks concepts, which are the storage capacity, patterns restoration and so on. From this point of view two results are of special interest.

First, notice that if $\tau \geq 3$ for the pattern with individual weight $r_1 = \tau$ the phase transition of the first kind disappears. We recall that when $\tau$ increases the critical value of the load parameter $\alpha_c^{(1)}$ also increases, the overlap of the local minimum with the pattern $m_c^{(1)}$ steadily decreases. Beginning from $\tau = 3$ the overlap in the critical point vanishes (see graphs on the left panels in Fig. 3). This means that the phase transition ceases to exist. As a rule disappearance of phase transition signifies substantial structural changes in the system.

Second, it is interesting how characteristics of local minima located near patterns with weight coefficients $r_\mu = 1$ ($\mu \geq 2$) depend on the value of $\tau$. We recall that at first increase in $\tau$ does not affect these local minima at all. While $\tau < 5.568$ neither the overlap of the local minimum with the pattern, nor its critical characteristics depend on $\tau$ (see Eq. (14)). It seems rather reasonable since the number of patterns with weight coefficients $r_\mu = 1$ is very large ($M \gg 1$). Therefore, as would be expected, the influence of only one pattern with the weight $\tau$ is negligible small. However, as soon as $\tau$ exceeds the critical value $\tau_c \approx 5.568$ (which is not so large) its influence on the huge number of local minima from the vicinities of patterns with weight coefficients $r_\mu = 1$ becomes rather noticeable (see Eqs. (15)). Let us emphasize that all the results are justified by computer simulations.

In conclusion we note that Hebbian connection matrix with different weight coefficients $r_\mu$ was also examined previously (see, for example [12], [13]). In [13] basic equations for this connection matrix were obtained. They coincide with our system of equations (4). However, then the authors simplified their equations and only the standard Hopfield model was analyzed. We see that even rather little differences in values of weights lead to new and nontrivial results. We hope to use the same approach for analysis of local minima of the quadratic functional (1) with an arbitrary connection matrix (see Introduction and papers [6]-[8]).

The work was supported by the program of Russian Academy of Sciences "Information technologies and analysis of complex systems" (project 1.7) and in part by Russian Basic Research Foundation (grant 09-07-00159).